\documentclass[prb,twocolumn,showpacs]{revtex4}
\usepackage[a4paper,hmargin=1in]{geometry}
\usepackage{amsmath,amssymb}
\usepackage{graphicx}
\usepackage{relsize}
\usepackage{bm}
\renewcommand{\vec}[1]{\bm{#1}}
\newcommand{\mat}[1]{\mathsf{#1}}
\newcommand{\D}{\text{d}}
\newcommand{\E}{\text{e}}
\newcommand{\Exp}[1]{\mathlarger{\E^{#1}}}%
\begin{document}
\title{Evaluation of Pressure Tensor in Constant-Volume Simulations of Hard and
  Soft Convex Bodies}
\author{Michael P. Allen}
\affiliation{Department of Physics and Centre for Scientific Computing \\
University of Warwick, Coventry CV4 7AL, UK}
\begin{abstract}
A method for calculating the pressure tensor in constant-volume Monte Carlo
simulations of convex bodies is presented. In contrast to other approaches, the
method requires only an isotropic scaling of the simulation box, and the
counting of simple geometric quantities characterizing overlapping
pairs. Non-sphericity presents no special difficulties. The result is expressed
as a sum of pairwise contributions, and can therefore be used to compute
pressure tensor profiles in a conventional way.
\end{abstract}
\maketitle
\section{Introduction}
The purpose of this brief paper is to explain in detail a procedure to calculate
the pressure tensor in constant-volume computer simulations of hard convex
bodies.  The term ``hard'' is used to denote particles whose interaction energy
is infinite when they overlap, and zero when they do not; this paper will also
consider ``soft'' convex bodies, meaning particles whose interaction energy
takes a finite, positive, constant value $\varepsilon$ when they overlap, and
zero when they do not. The restriction to convex bodies is a simplification to
ensure that an overlap between a pair of molecules may result when the distance
between the molecular centres is reduced, but not when it is increased, while
keeping all orientations fixed; extending the results to non-convex bodies is
relatively straightforward but requires care. Calculation of the components of
the pressure tensor, as opposed to the scalar pressure, is an important route to
the surface tension of interfaces.

The motivation for this work is the recent paper by \citet{gloor.gj:2005.a}
which, as well as providing a near-comprehensive review of simulation techniques
in this area, proposes a method for surface tension calculation based on
perturbations in the cross-sectional area of a system containing the planar
interface of interest. It is claimed that a key advantage of this ``test-area''
method for discontinuous potentials is the avoidance of the determination of
delta-functions ``which are impractical to evaluate, particularly in the case of
non-spherical molecules''. The present paper demonstrates that this is not the
case: that the pressure tensor components for such systems, and hence the
surface tension, may be calculated by a simple extension of the standard
approach of counting overlaps produced by an isotropic volume scaling. In fact,
this method has already been used to determine the surface tension of the
isotropic-nematic interface of hard ellipsoids,\cite{mcdonald.aj:2000.a} 
although the full details of the method were not explained in that paper.

Microscopic expressions for the pressure in a fluid are to be found in the
standard references.\citep{hansen.jp:1986.a,gray.cg:1984.a} Separate
$P=P^\text{id}+P^\text{ex}$ where $P^\text{id}=\rho k_\text{B}T$ is the ideal
gas contribution, and focus on the excess part, $P^\text{ex}$.  The usual
mechanical route to the pressure starts with the volume derivative of the excess
Helmholtz free energy $ F^\text{ex}$
\begin{equation}
\frac{P^\text{ex}}{k_\text{B}T}
= -\frac{1}{k_\text{B}T}\frac{\partial F^\text{ex}}{\partial V}
= \frac{\partial \ln Q^\text{ex}}{\partial V}
= \frac{1}{Q^\text{ex}}\frac{\partial Q^\text{ex}}{\partial V} \:,
\label{eqn:pex1}
\end{equation}
where $T$ is the temperature and $k_\text{B}$ Boltzmann's constant.  The excess
partition function in the canonical ensemble, $Q^\text{ex}$, is written
\begin{align}
Q^\text{ex} &= V^{-N} \int \D\vec{r}_1\cdots\int\D\vec{r}_N \;
\Exp{-U/k_\text{B}T}
\nonumber \\
&= \int \D\vec{s}_1\cdots\int \D\vec{s}_N \; \Exp{-U/k_\text{B}T} \;,
\label{eqn:qex}
\end{align}
where $U$ is the total potential energy.  The $N$ particle centre-of-mass
coordinates are denoted $\vec{r}_j$, and scaled coordinates $\vec{s}_j$ are
defined by $\vec{r}_j=L\vec{s}_j$, assuming a cubic box of side $L$ and volume
$V=L^3$. Particles may have orientational degrees of freedom; for example the
orientation of a uniaxial molecule is specified by a unit vector
$\hat{\vec{u}}_j$ directed along the symmetry axis.  However, for simplicity
here, and in most of what follows, the orientational dependence and the
integrations over orientational degrees of freedom are not explicitly written.
In the following, also for simplicity, pairwise-additive potentials
$U=\sum_{i<j} u_{ij}$ with
$u_{ij}=u(\vec{r}_{ij},\hat{\vec{u}}_i,\hat{\vec{u}}_j)$, where $\vec{r}_{ij}=
\vec{r}_{i}- \vec{r}_{j}$, are assumed.
For continuous potentials $u_{ij}$, the volume differentiation within the
integral of eqn~\eqref{eqn:qex} is readily carried out yielding the well-known
expression
\begin{align}
\frac{P^\text{ex}}{k_\text{B}T}
&= -\frac{1}{3Vk_\text{B}T}
\left\langle \sum_{i<j} 
\left(\frac{\partial u_{ij}}{\partial \vec{r}_{ij}}\right)\cdot \vec{r}_{ij} \right\rangle
\nonumber \\
&= -\frac{1}{3Vk_\text{B}T}
\left\langle \sum_{i<j} 
\left(\frac{\partial u_{ij}}{\partial r_{ij}}\right) r_{ij} \right\rangle \;.
\label{eqn:pex2}
\end{align}
Any orientations (for example $\hat{\vec{u}}_i$, $\hat{\vec{u}}_j$, and
$\hat{\vec{r}}_{ij}=\vec{r}_{ij}/r_{ij}$) are held fixed in taking the
derivative in the right-most expression in eqn~\eqref{eqn:pex2}.
\section{Pressure for Discontinuous Potentials}
For discontinuous potentials, $\partial Q^\text{ex}/\partial V$ may be estimated
by finite differences. To highlight the connection with the method for
determining the full pressure tensor, a derivation is presented here; this
follows \citet{eppenga.r:1984.a,perram.jw:1984.a}, and the basic ideas go back
to \citet{vieillard-baron:1972.a}, although the detailed approach and notation
used here are different.  Consider a volume change $V\rightarrow V-\Delta V$.
\begin{align*}
\frac{P^\text{ex}}{k_\text{B}T}
&=
\frac{1}{Q^\text{ex}}\frac{\partial Q^\text{ex}}{\partial V}
\\
&= \lim_{\Delta V\rightarrow 0_+} \frac{1}{\Delta V} 
\frac{Q^\text{ex}(V)-Q^\text{ex}(V-\Delta V)}{Q^\text{ex}(V)}
\\
&=  \lim_{\Delta V\rightarrow 0_+}
\frac{1}{\Delta V} \left[
1 - \left\langle \Exp{-\Delta U/k_\text{B}T} \right\rangle
\right] \;.
\end{align*}
The ensemble average is conducted in the unperturbed system and $\Delta
U=\sum_{i<j} \Delta u_{ij}$ is a sum of energy changes for each pair, arising
from overlaps which are freshly generated by the volume change.  Hence the
Boltzmann factor is expressed as a product of terms. The expression is
simplified by the assumption that the different pair overlaps are uncorrelated.
This is reasonable at small $\Delta V$, since the number of overlaps may be made
vanishingly small.
\begin{align*}
\frac{P^\text{ex}}{k_\text{B}T}
&= \lim_{\Delta V\rightarrow 0_+}
\frac{1}{\Delta V} \left[
1 - \left\langle \prod_{i<j} \Exp{-\Delta u_{ij}/k_\text{B}T} \right\rangle
\right]
\\
&= \lim_{\Delta V\rightarrow 0_+}
\frac{1}{\Delta V} \left[
1 - \prod_{i<j} \left\langle \Exp{-\Delta u_{ij}/k_\text{B}T} \right\rangle
\right] \;.
\end{align*}
All pairs are now equivalent in this expression.  For hard particles, each
exponential is either 0 (signalling a newly-generated overlap) or 1 (when the
volume scaling does not cause an overlap), and the average has the
interpretation of a probability that any given pair will \emph{not} overlap. For
soft particles, every newly-generated overlap contributes
$\Exp{-\varepsilon/k_\text{B}T}$, while pairs whose overlap status remains
unchanged contribute 1. In either case, it is convenient to define
\begin{align}
\phi_{ij} &\equiv 1- \Exp{-\Delta u_{ij}/k_\text{B}T} 
\nonumber \\ &=
\begin{cases}
1- \Exp{-\varepsilon/k_\text{B}T} & \text{newly generated overlap} \\
0 & \text{otherwise}
\end{cases}
\label{eqn::phidef}
\end{align}
\begin{equation}
\mathcal{P}_{ij}^\text{overlap}
\equiv \left\langle \phi_{ij} \right\rangle
= 1-\left\langle \Exp{-\Delta u_{ij}/k_\text{B}T} \right\rangle \;.
\label{eqn::Pdef}
\end{equation}
$\mathcal{P}_{ij}^\text{overlap}$ will be loosely termed an ``overlap
probability''; for small $\Delta V$, it is small compared with 1, simply because
most pairs will have $\Delta u_{ij}=0$. The indices $i$ and $j$ on
$\mathcal{P}_{ij}^\text{overlap}$ are actually redundant.  The product of terms
is expanded to first order in the overlap probabilities:
\begin{align*}
\frac{P^\text{ex}}{k_\text{B}T}
&= \lim_{\Delta V\rightarrow 0_+}
\frac{1 - \prod_{i<j} \bigl( 1-\mathcal{P}_{ij}^\text{overlap}
  \bigr)}%
{\Delta V}
\\
&= \lim_{\Delta V\rightarrow 0_+}
\frac{\sum_{i<j} \mathcal{P}_{ij}^\text{overlap}}%
{\Delta V} 
\\
&= \lim_{\Delta V\rightarrow 0_+}
\frac{\left\langle\sum_{i<j} \phi_{ij}\right\rangle}%
{\Delta V} 
\equiv \lim_{\Delta V\rightarrow 0_+}
\frac{\left\langle \Phi^\text{overlap} \right\rangle}{\Delta V}  \;.
\end{align*}
For hard particles, $\Phi^\text{overlap}=N^\text{overlap}$, the
number of overlaps in a configuration generated by the volume scaling, arising
from summing the probabilities over all pairs. In the more general case of soft
particles
\begin{align*}
\Phi^\text{overlap} &= \sum_{i<j} \phi_{ij} = 
\sum_{i<j} 1- \Exp{-\Delta u_{ij}/k_\text{B}T}
\\
&= N^\text{overlap} \bigl( 1 - \Exp{-\varepsilon/k_\text{B}T} \bigr) \;,
\end{align*}
re-emphasizing that $N^\text{overlap}$ counts the number of \emph{new} overlaps,
since the original configuration may already contain overlaps if $\varepsilon$
is finite, and these do not contribute.

Suppose now that the volume reduction comes from scaling the box lengths by a
factor $(1-\epsilon)$, i.e.\ $L\rightarrow L-\Delta L$ where $\Delta
L/L=\epsilon$, then $\Delta V/V\approx 3\epsilon$ and
\begin{align}
\frac{P^\text{ex}}{k_\text{B}T}
&= \lim_{\Delta V\rightarrow 0_+}
\frac{\left\langle \Phi^\text{overlap} \right\rangle}{\Delta V} 
\nonumber \\ 
&= \lim_{\epsilon\rightarrow 0_+}
\frac{\left\langle \Phi^\text{overlap} \right\rangle}{3\epsilon V}
= \lim_{\epsilon\rightarrow 0_+} 
\frac{\left\langle \sum_{i<j} \phi_{ij} \right\rangle}{3\epsilon V} \;.
\label{eqn:pex3}
\end{align}
Thus the pressure may be estimated by essentially counting the overlaps that
result from a small isotropic volume scaling. The same expression (to leading
order in $\epsilon$) results if the overlaps result from scaling all
linear particle dimensions by the factor $1+\epsilon$ instead of reducing the
volume.

The expression of \citet{perram.jw:1984.a} for
convex spheroids is equivalent. It may be written
\begin{equation}
\frac{P^\text{ex}}{k_\text{B}T} = 
 \lim_{\epsilon\rightarrow 0_+} 
\frac{\left\langle\sum_{i<j}' F_{ij} \right\rangle}{3\epsilon V} \;,
\label{eqn:pw}
\end{equation}
with overlap function $F_{ij}$ defined such that
\begin{alignat*}{2}
F_{ij} &>1 && \text{, for non-overlapping $i$ \& $j$,} \\
F_{ij} &=1 && \text{, for $i$ \& $j$ in contact,} \\
F_{ij} &<1 && \text{, for overlapping  $i$ \& $j$.} 
\end{alignat*}
The sum in eqn~\eqref{eqn:pw} counts all distinct pairs for which
$1<F_{ij}<(1+\epsilon)^2\approx 1+2\epsilon$. The value $F_{ij}$ within the sum
can be replaced by 1. The overlap function defined by \citet{perram.jw:1984.a}
has the scaling behaviour $F_{ij}=r^2
f_{ij}(\hat{\vec{r}}_{ij},\vec{u}_i,\vec{u}_j)$ so counting pairs for which
$1<F_{ij}<\approx (1+\epsilon)^2$ is equivalent to counting the overlaps
generated by scaling all particle linear dimensions by $1+\epsilon$ as above.

\section{Pressure Tensor for Discontinuous Potentials}
Turning now to the pressure tensor, the ideal part is
$P_{\alpha\beta}^\text{id}/k_\text{B}T = \rho\delta_{\alpha\beta}$ and the
excess part may be expressed \citep{hansen.jp:1986.a}
\begin{gather*}
\frac{P_{\alpha\beta}^\text{ex}}{k_\text{B}T}
= -\frac{1}{Vk_\text{B}T}\left\langle 
\sum_{i<j} \frac{\partial u(r_{ij})}{\partial r_{ij\alpha}} r_{ij\beta} 
\right\rangle\;, 
\: \alpha,\beta = x,y,z
\\
\text{or}\quad
\frac{\mat{P}^\text{ex}}{k_\text{B}T}
= -\frac{1}{Vk_\text{B}T}\left\langle 
\sum_{i<j} \frac{\partial u(r_{ij})}{\partial \vec{r}_{ij}} \otimes \vec{r}_{ij} 
\right\rangle 
\end{gather*}
where $\otimes$ represents an outer product. To estimate this tensor in the case
of discontinuous potentials, a more detailed consideration of pair geometry is
required.
\begin{figure}
% Figure prepared from geometry2.fig -> geometry2.eps
% Then processed with geometry2.tex using psfrag
% Finally converted .ps -> .eps
\includegraphics[width=0.5\textwidth,clip]{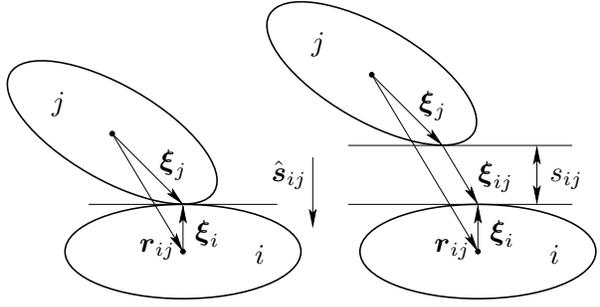}
\caption{\label{fig:1}%
  Schematic geometry of two convex particles in contact (left) and near contact
  (right). The vectors $\vec{\xi}_i$ and $\vec{\xi}_j$, as well as the particle
  orientations, are kept fixed as the centre-centre vector $\vec{r}_{ij}$ is
  scaled uniformly. The separation distance is defined between the two tangent
  planes. After Ref.~\protect\onlinecite{allen.mp:1993.h}.}
\end{figure}
When the particles are in contact, there is a common tangent plane at the point
of contact; define a unit vector $\hat{\vec{s}}_{ij}$ normal to this plane, from
$j$ to $i$, as shown in Fig.~\ref{fig:1}. For small displacements of the
centre-centre vector $\vec{r}_{ij}$, keeping all angles, and the vectors
$\vec{\xi}_i$ and $\vec{\xi}_j$ of the contact points relative to the molecular
centres, fixed, the contact points and their associated tangent planes separate.
Define the separation vector $\vec{\xi}_{ij}$, and the perpendicular separation
between the tangent planes $s_{ij}=\hat{\vec{s}}_{ij}\cdot\vec{\xi}_{ij}$, as
shown in the figure.  Then $s_{ij}$ takes positive values if the particles do
not overlap, and negative values if they do overlap. Hence the step function
$\theta(s_{ij})$ takes the value 0 (overlap) or 1 (no overlap), and a standard
trick \cite{boublik.t:1974.a,nezbeda.i:1977.a,aviram.i:1977.a,boublik.t:1984.a}
may be used to estimate the gradient of the potential. Write
\begin{equation*}
\Exp{-u_{ij}/k_\text{B}T} = \theta(s_{ij}) 
+ \bigl(1-\theta(s_{ij})\bigr)\Exp{-\varepsilon/k_\text{B}T} \;,
\end{equation*}
(the second term is zero for hard particles).  Take the gradient of both sides
and rearrange
\begin{gather*}
 - \frac{1}{k_\text{B}T} \frac{\partial u(r_{ij})}{\partial \vec{r}_{ij}}
  =  \left(\frac{\partial s_{ij}}{\partial\vec{r}_{ij}}\right)
  \delta(s_{ij}-0_+) \times
\\
\times
 \bigl(1-\Exp{-\varepsilon/k_\text{B}T}\bigr)\Exp{+u_{ij}/k_\text{B}T}
\end{gather*}
Note that the delta function may be taken to act at a separation infinitesimally
outside contact, and so the exponential $\Exp{+u_{ij}/k_\text{B}T}$ may be
replaced by unity.  Also, $\left(\dfrac{\partial
    s_{ij}}{\partial\vec{r}_{ij}}\right) =\hat{\vec{s}}_{ij}$, so the expression
simplifies:
\begin{equation*}
 - \frac{1}{k_\text{B}T} \frac{\partial u(r_{ij})}{\partial \vec{r}_{ij}}
  =  \hat{\vec{s}}_{ij} \delta(s_{ij}-0_+)
  \bigl(1-\Exp{-\varepsilon/k_\text{B}T}\bigr)
\:.
\end{equation*}
This is a straightforward generalization of the expression of
\citet{boublik.t:1974.a} to include the case of soft particles. The connection
to the pair distribution functions of hard convex body fluids at contact is set
out elsewhere.\cite{allen.mp:1993.h}  For the present purposes, the delta function
is approximated \cite{trokhymchuk.a:1999.a} by the term in braces below:
\begin{gather*}
 - \frac{1}{k_\text{B}T} \frac{\partial u(r_{ij})}{\partial \vec{r}_{ij}}
 = 
\hat{\vec{s}}_{ij} \times
\\ \times
\left\{\lim_{\Delta s\rightarrow 0_+} 
\frac{\theta(s_{ij})-\theta(s_{ij}-\Delta s)}{\Delta s}
\right\} \bigl(1-\Exp{-\varepsilon/k_\text{B}T}\bigr)
\;.
\end{gather*}
If the change results from a scaling $\vec{r}_{ij}\rightarrow
\vec{r}_{ij}-\Delta\vec{r}_{ij}$ with $\Delta\vec{r}_{ij}=\epsilon\vec{r}_{ij}$,
then $\Delta s = \hat{\vec{s}}_{ij}\cdot\Delta\vec{s}_{ij} =
\hat{\vec{s}}_{ij}\cdot\Delta\vec{r}_{ij} =
\epsilon\hat{\vec{s}}_{ij}\cdot\vec{r}_{ij} $.  The remaining terms may be
recognized as $\phi_{ij}$, defined in eqn~\eqref{eqn::phidef}.  The result
becomes
\begin{gather}
 - \frac{1}{k_\text{B}T} \frac{\partial u(r_{ij})}{\partial \vec{r}_{ij}}
  = \lim_{\epsilon\rightarrow 0_+} 
  \frac{\phi_{ij}\hat{\vec{s}}_{ij}}{\epsilon\hat{\vec{s}}_{ij}\cdot\vec{r}_{ij}}
\nonumber
\\
\Rightarrow\quad
\frac{\mat{P}^\text{ex}}{k_\text{B}T}  = \lim_{\epsilon\rightarrow 0_+} 
\frac{1}{\epsilon V} \left\langle \sum_{i<j} \phi_{ij}
\frac{\hat{\vec{s}}_{ij}\otimes\vec{r}_{ij}}{\hat{\vec{s}}_{ij}\cdot\vec{r}_{ij}} 
\right\rangle
\:.
\label{eqn:pex4}
\end{gather}
This is the main result: pressure tensor components in constant-volume Monte
Carlo simulations of general convex bodies can be calculated almost as simply as
the scalar pressure, by summing over pairs which would overlap as a result of an
isotropic scaling of coordinates. The term $\phi_{ij}$ acts to select such pairs
(and for soft particles, it also incorporates the appropriate energy-dependent
scale factor); the only additional requirement is to compute the surface normal
for the near-contact pairs, which is almost always straightforwardly done. The
above derivation follows closely that of \citet{trokhymchuk.a:1999.a}, although
that paper explicitly considered only the spherically symmetric case. Taking one
third of the trace of eqn~\eqref{eqn:pex4} regenerates eqn~\eqref{eqn:pex3}.

This expression is consistent with the form of pressure tensor calculated in
collisional molecular dynamics simulations
\begin{equation*}
\mat{P}^\text{ex} = \frac{1}{Vt} \sum_\text{colls} 
\Delta p_{ij} \hat{\vec{s}}_{ij}\otimes\vec{r}_{ij}
\end{equation*}
which is a sum over collisions occurring in time $t$, where the colliding
particles are labelled $i$ and $j$, and the impulsive momentum transfer of magnitude
$\Delta p_{ij}$ is directed along the common surface normal
$\hat{\vec{s}}_{ij}$.  For continuous potentials, the analogous formula is
proportional to $\vec{f}_{ij}\otimes\vec{r}_{ij}$ where $\vec{f}_{ij}$ is the
pair force. Indeed, eqn~\eqref{eqn:pex4} may be written down almost immediately,
knowing that pair contributions must be proportional to
$\hat{\vec{s}}_{ij}\otimes \vec{r}_{ij}$, and that the isotropic pressure is
given by eqn~\eqref{eqn:pex3}.

To evaluate the surface tension $\gamma$ of a planar interface normal to the $z$
direction, the difference in components $\Delta P(z) \equiv
P_{zz}-\frac{1}{2}(P_{xx}(z)+P_{yy}(z))$ is calculated as a function of $z$ and
integrated across the interface. The integrals may be computed implicitly,
yielding $\gamma$ as a single number, but the above formalism makes explicit the
contributions of each pair to the total for each component of $\mat{P}$, and
therefore the definition of a local pressure tensor $\mat{P}(z)$ via the
Irving-Kirkwood or Harasima convention
\cite{irving.jh:1950.a,harasima.a:1958.a,schofield.p:1982.a} is handled in
exactly the same way as for continuous potentials. The integration of $\Delta
P(z)$ is then performed numerically. As emphasized by \citet{holcomb.cd:1993.a}
and \citet{trokhymchuk.a:1999.a}, monitoring pressure tensor profiles in
inhomogeneous systems is an important test of equilibrium, as well as providing
useful information about interface structure. This aspect is missing from the
test-area method as described in Ref.~\onlinecite{gloor.gj:2005.a}, although it
could probably be incorporated.
\section{Conclusions}
This paper has demonstrated that all components of the pressure tensor may be
straightforwardly computed in constant-volume simulations by counting the pair
overlaps resulting from isotropic scaling of the simulation box, and including a
simple geometrical tensor formed from the surface normal at contact and the
centre-centre vector. Non-spherical particles with discontinuous interaction
potentials present no great difficulties and special anisotropic scaling moves
are not required.  The method makes it easy to resolve the pressure tensor
profile, for use in surface tension calculations.

No explicit comparisons of techniques are performed here, but it should be noted
that equation \eqref{eqn:pex4} has been used to calculate pressure tensor
profiles, and the surface tension, of the equilibrium isotropic-nematic
interface in the hard ellipsoid fluid,\cite{mcdonald.aj:2000.a} and comparisons
made there with results for a closely-related continuous potential model. There
is a trade-off between minimizing statistical errors (favoured by large numbers
of overlaps) and systematic errors (favoured by small volume changes), in the
choice of the scaling parameter $\epsilon$, so usually it will be necessary to
examine several values.  As has been emphasized, the method is essentially
equivalent to a numerical estimate of the (orientationally-averaged) pair
distribution function at contact, and so the same caveats apply as to the
calculation of that quantity:\cite{allen.mp:1989.j} near contact, the
function may vary quite sharply, so the extrapolation to small $\epsilon$ may
require care. These observations apply equally to the test-area method, and all
other finite perturbation methods of this kind.
\begin{acknowledgments}
  This research has been supported by the Engineering and Physical Sciences
  Research Council, and by the Alexander von Humboldt foundation while the
  author was on Study Leave at the Johannes Gutenberg University, Mainz.
  Conversations with Tanja Schilling and David Cheung are gratefully
  acknowledged. Computing facilities were provided by the Centre for Scientific
  Computing, University of Warwick.
\end{acknowledgments}
\bibliography{journals,main}
\end{document}